\documentclass[aps,prl,superscriptaddress,floatfix,onecolumn, notitlepage, preprint,nobibnotes]{revtex4-2}
\usepackage{amsmath,amssymb,amsfonts}
\usepackage{hyperref}
\usepackage{graphicx}
\usepackage{ulem}
\usepackage[table,x11names]{xcolor}
\usepackage{textcomp, gensymb}
\usepackage{bm}
\usepackage{physics}
\usepackage{colortbl}
\usepackage{booktabs}
\usepackage[left]{lineno}

\makeatletter

    \def\CT@@do@color{%
      \global\let\CT@do@color\relax
            \@tempdima\wd\z@
            \advance\@tempdima\@tempdimb
            \advance\@tempdima\@tempdimc
    \advance\@tempdimb\tabcolsep
    \advance\@tempdimc\tabcolsep
    \advance\@tempdima2\tabcolsep
            \kern-\@tempdimb
            \leaders\vrule
                    \hskip\@tempdima\@plus  1fill
            \kern-\@tempdimc
            \hskip-\wd\z@ \@plus -1fill }
    \makeatother

\makeatletter
\def\blfootnote{\xdef\@thefnmark{}\@footnotetext}
\makeatother

\makeatletter
\newcommand{\fmarki}{*}
\newcommand{\fmarkii}{\ensuremath{\dagger}}
\newcommand{\fmarkiii}{\ensuremath{\ddagger}}
                
\def\@fnsymbol#1{{\ifcase#1\or \fmarki\or \fmarkii\or \fmarkiii \else\@ctrerr\fi}}
\makeatother
\renewcommand{\fmarki}{\ensuremath{\dagger}}
\renewcommand{\fmarkii}{\ensuremath{\ddagger}}
\renewcommand{\fmarkiii}{\ensuremath{\mathsection}}

\begin{document}

\include{main_figure_definitions}

\title{Large Exciton Binding Energy in the Bulk van der Waals Magnet CrSBr}

\author{Shane Smolenski}
\affiliation{Department of Physics, University of Michigan, Ann Arbor, MI 48109, USA}

\author{Ming Wen}
\affiliation{Department of Chemistry, University of Michigan, Ann Arbor, MI 48109, USA}

\author{Qiuyang Li}
\affiliation{Department of Physics, University of Michigan, Ann Arbor, MI 48109, USA}

\author{Eoghan Downey}
\affiliation{Department of Physics, University of Michigan, Ann Arbor, MI 48109, USA}

\author{Adam Alfrey}
\affiliation{Applied Physics Program, University of Michigan, Ann Arbor, MI 48109, USA}

\author{Wenhao Liu}
\affiliation{Department of Physics, University of Texas at Dallas, Richardson, TX 75080, USA}

\author{Aswin L. N. Kondusamy}
\affiliation{Department of Materials Science and Engineering, University of Texas at Dallas, Richardson, TX 75080, USA}

\author{Aaron Bostwick}
\affiliation{Advanced Light Source, Lawrence Berkeley National Laboratory, Berkeley, CA 94720, USA}

\author{Chris Jozwiak}
\affiliation{Advanced Light Source, Lawrence Berkeley National Laboratory, Berkeley, CA 94720, USA}

\author{Eli Rotenberg}
\affiliation{Advanced Light Source, Lawrence Berkeley National Laboratory, Berkeley, CA 94720, USA}

\author{Liuyan Zhao}
\affiliation{Department of Physics, University of Michigan, Ann Arbor, MI 48109, USA}

\author{Hui Deng}
\affiliation{Department of Physics, University of Michigan, Ann Arbor, MI 48109, USA}

\author{Bing Lv}
\affiliation{Department of Physics, University of Texas at Dallas, Richardson, TX 75080, USA}

\author{Dominika Zgid}
\affiliation{Department of Chemistry, University of Michigan, Ann Arbor, MI 48109, USA}

\author{Emanuel Gull}
\affiliation{Department of Physics, University of Michigan, Ann Arbor, MI 48109, USA}

\author{Na Hyun Jo}
\altaffiliation{nhjo@umich.edu}
\affiliation{Department of Physics, University of Michigan, Ann Arbor, MI 48109, USA}

\date{\today}
\maketitle

\def\kill #1{\sout{#1}}
\def\nhj #1{\textcolor{blue}{#1}}   
\def\nhjcomment #1{\nhj{[nhj: #1]}} 
\def\sas #1{\textcolor{red}{#1}}

\section{Abstract}
Excitons, bound electron-hole pairs, influence the optical properties in strongly interacting solid state systems. Excitons and their associated many-body physics are typically most stable and pronounced in monolayer materials. Bulk systems with large exciton binding energies, on the other hand, are rare and the mechanisms driving their stability are still relatively unexplored. Here, we report an exceptionally large exciton binding energy in single crystals of the bulk van der Waals antiferromagnet CrSBr. Utilizing state-of-the-art angle-resolved photoemission spectroscopy and self-consistent ab-initio \textit{GW} calculations, we present direct spectroscopic evidence that robust electronic and structural anisotropy can significantly amplify the exciton binding energy within bulk crystals. Furthermore, the application of a vertical electric field enables broad tunability of the optical and electronic properties. Our results indicate that CrSBr is a promising material for the study of the role of anisotropy in strongly interacting bulk systems and for the development of exciton-based optoelectronics.  

\section{Main}
Excitons are energetically favorable bound states of excited electrons and holes with an energy less than the single-particle (electronic) band gap~\cite{KOCH2006}. This energy difference between the exciton energy and the electronic gap is the exciton binding energy. Large exciton binding energies, resulting from increased Coulomb interactions, stabilize excitons up to room temperature~\cite{MUELLER2018} and enhance the associated emergent phenomena, such as exciton condensates~\cite{WANG2019}, excitonic insulators~\cite{JIA2022,CHEN2023,CHEN2023_NATURE}, and charged excitons~\cite{ROSS2013}. In addition to stabilization, large binding energies lead to enhanced absorption, sharp spectral emission, and electrical tunability~\cite{ROSS2014}, all of which are promising properties for next-generation optoelectronic applications, including photovoltaics~\cite{BERNARDI2013} and single photon emitters~\cite{KOPERSKI2015}. These large exciton binding energies are most prevalent in monolayer materials, namely monolayer transition metal dichalcogenides (TMDs) where binding energies exceed 500\,meV, because of reduced dielectric screening and geometrical confinement~\cite{MUELLER2018,DVORAK2013,POKUTNYI2020,ZHANG2018,JUNG2022}. Bulk materials, with significantly stronger screening environments, typically host excitons with binding energies orders of magnitude lower~\cite{ZHANG2018} yet would provide a platform on which to explore the role of dimensionality in many-body interactions. In theory, bulk materials with strong Coulomb interactions encouraged by anisotropy and/or reduced screening could overcome this limitation~\cite{SHARMA2018,BALANDIN2022}. Particularly, van der Waals magnets, in which anisotropy is necessary to stabilize magnetic order down to the 2D limit and which host exotic quasiparticle interactions~\cite{MERMIN1966, GIBERTINI2019, JIANG2021,YOU2019,JIANG2018}, could thus provide promising materials in which to detect tightly bound bulk excitons and explore their rich many-body physics.

\begin{figure} [ht]
    \includegraphics[width = 3.5 in]{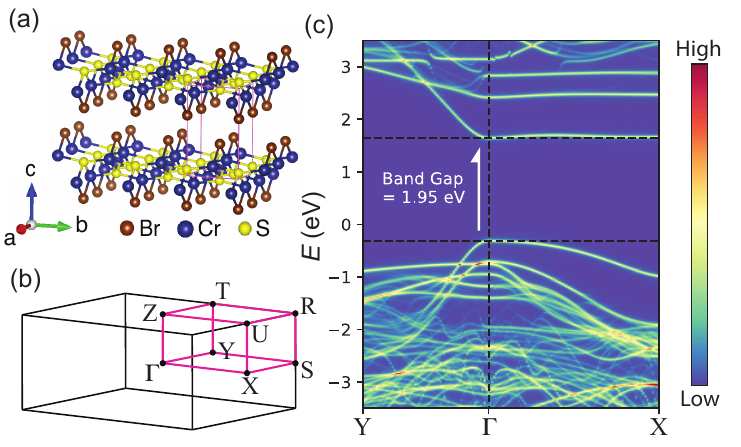}%
    \caption{\textbf{Crystal and Electronic Structure of CrSBr.} 
    (a)	Three-dimensional representation of the real-space crystal structure of CrSBr, created with the VESTA software package~\cite{MOMMA2011}. The pink rectangular prism represents the unit cell. (b) The corresponding Brillouin zone of CrSBr with high symmetry points labeled. (c) Self-consistent \textit{GW} calculations of the electronic band structure of monolayer CrSBr along high symmetry directions in the $\Gamma$XSY plane.
    \label{fig:Fig1}}
\end{figure}

Among 2D magnets, CrSBr stands out as a highly promising candidate due to its strong optical response and pronounced anisotropy. CrSBr is a van der Waals antiferromagnet that crystallizes in the orthorhombic space group Pmmn (space group number 59) with lattice parameters $a\,=\,3.50\,\text{\AA}$, $b\,=\,4.75\,\text{\AA}$, and $c\,=\,7.94\,\text{\AA}$ near $T\,=\,200$\,K~\cite{GOSER1990,LIU2022}. Its structure is comprised of van der Waals layers in the \textit{ab} plane with each layer consisting of two buckled CrS sheets capped by Br (Fig.~\ref{fig:Fig1} (a)). The corresponding Brillouin zone is shown in Fig.~\ref{fig:Fig1} (b). Below the Néel temperature $T_N\,=\,132$\,K, CrSBr develops \textit{A}-type antiferromagnetic (AFM) order where the Cr magnetic moments couple ferromagnetically within the \textit{ab} plane but antiferromagnetically in the stacking direction with the \textit{b} axis being the easy axis~\cite{GOSER1990,TELFORD2020}. The monolayer has also been proven to remain stable in air and exhibits ferromagnetic ordering within the plane below 146\,K\,\cite{LEE2021,LIU2022}. Moreover, an extraordinary phase transition, which is indicative of weak interlayer coupling, has been revealed in bulk CrSBr where the surface transitions into the AFM state at 140\,K while the bulk transitions at 132\,K~\cite{GUO2023}. Importantly, CrSBr exhibits a large optical response arising from Wannier excitons with binding energies predicted to be between 500-900\,meV for the monolayer~\cite{WILSON2021,KLEIN2023}. Correspondingly, a large exciton-photon coupling strength of $\sim20$\,meV per monolayer was measured in 2D CrSBr polaritons~\cite{LI2023}. Coupling between charge, spin, and lattice further demonstrate the presence of rich quasi-particle interactions~\cite{WILSON2021,KLEIN2023,LIN2024,CENKER2022,LI2023}. Alongside structural and magnetic anisotropy, calculations and electrical transport measurements suggest that the electronic structure exhibits strong electronic anisotropy in the form of a quasi-one-dimensional (quasi-1D) conduction band~\cite{KLEIN2023,WU2022}. This is also well captured by our self-consistent \textit{GW} (sc\textit{GW}) calculation on monolayer CrSBr shown in Fig.~\ref{fig:Fig1} (c). From these optical and computational results, it has been suggested that CrSBr hosts tightly-bound quasi-1D exctions~\cite{WILSON2021,KLEIN2023,WANG2023}. However, direct experimental observation of these remarkable excitonic properties and their underlying mechanisms has not yet been realized.

In this study, we report the direct measurement of an exceptionally large exciton binding energy in single crystals of CrSBr through angle-resolved photoemission spectroscopy (ARPES). We attribute the large binding energy to strong charge localization due to electronic and structural anisotropy as well as enhanced Coulomb interactions from diminished dielectric screening, confirming the inferences from optical and computational results. To our knowledge, this is the first direct measurement of a large exciton binding energy in a 2D magnet. We also demonstrate that the introduction of additional carriers as well as the application of a vertical electric field can both reduce the band gap and thus tune the electronic and optical properties. Through our study, we shed light on the role of anisotropy in the stabilization and tunability of many-body interactions in bulk excitonic materials.

\begin{figure} [ht]
    \includegraphics[width = 6.3 in]{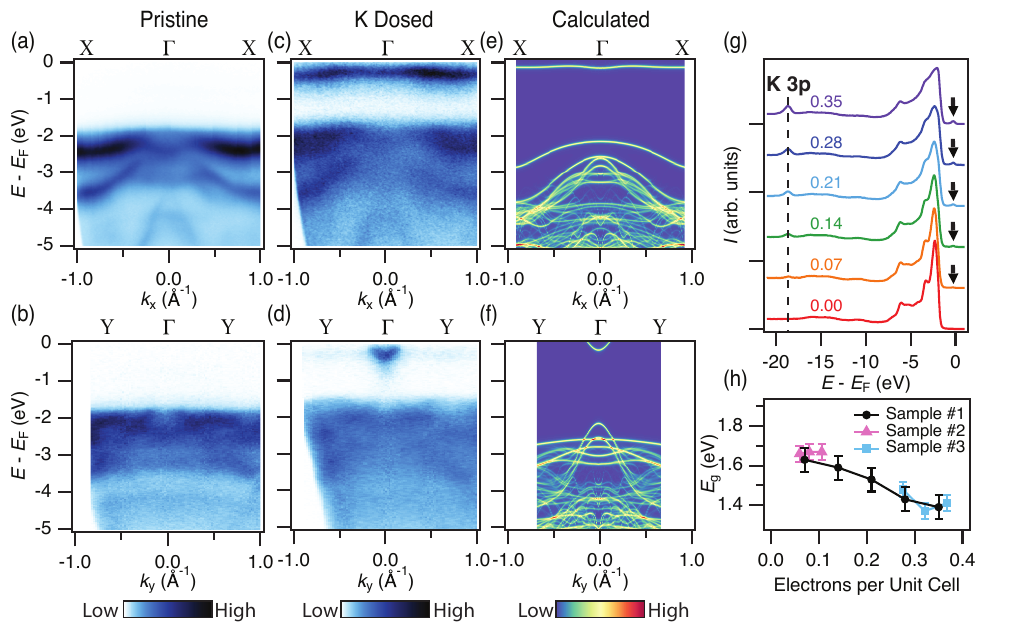}%
    \caption{\textbf{Electronic Stucture of CrSBr.} 
    (a)-(d) Experimental band dispersion along the $\Gamma-$X and $\Gamma-$Y high symmetry lines for (a,b) pristine and (c,d) highly dosed (0.35 electrons per unit cell) bulk CrSBr at $T\,=\,192$\,K (PM regime). The emergence of the conduction band and upward energy shift of the valence band upon dosing are evident. (e)-(f) sc\textit{GW} calculations of monolayer CrSBr along the (e) $\Gamma-$X and (f) $\Gamma-$Y. (g) Energy dispersion curves (EDCs) as a function of K dosing detail the gradual emergence of conduction band as a function of K dosing. The numbers above each spectrum corresponds to the dosing level in units of electrons per unit cell. The downward shift in energy of the conduction band along with the upward shift of the valence bands results in a reduction of the band gap as a function of K dosing (h).
    \label{fig:Fig2}}
\end{figure}

To evaluate the exciton binding energy, we must compare the electronic  gap (the energy necessary to separately create an electron and a hole) with the optical gap (the energy necessary to create a bound electron-hole pair). We first explored the electronic gap by performing ARPES measurements on single crystals of CrSBr in the paramagnetic (PM) phase (Fig.~\ref{fig:Fig2}). Figures~\ref{fig:Fig2} (a) and (b) show the band dispersion of pristine CrSBr along the high symmetry directions $\Gamma-$X and $\Gamma-$Y, respectively. Note that high symmetry points are designated based on photon energy scans and iso-energy plots at high binding energies (see Supplementary Information (SI) Fig.~\ref{fig:SM1}). The shape of the band structure along the high symmetry directions is in strong agreement with previously published ARPES data~\cite{BIANCHI2023BULK} and our sc\textit{GW} calculations (Figs.~\ref{fig:Fig2} (e), (f)). All of these results corroborate the insulating nature of the material, as evidenced by the lack of intensity at the Fermi energy $E_{F}$. It should be noted that while our calculations predict a valence band maximum (VBM) at $\Gamma$, we do not immediately observe this in the ARPES data due to matrix element effects in the first Brillouin zone~\cite{BIANCHI2023BULK}. However, the maximum at $\Gamma$ can be observed in the second Brillouin zone and we thus determine the VBM to be at an energy of 1.84~eV below $E_F$ (SI Fig.~\ref{fig:SM1}). Since we do not observe the conduction band in our pristine samples, this implies that the electronic gap is greater than 1.84\,eV. This measure of the lower bound on the electronic gap is slightly larger than the gap of $1.5~\pm~0.2$\,eV found with scanning tunneling spectroscopy (STS)~\cite{TELFORD2020,KLEIN2023}. Additionally, previous bulk ARPES measurements found the VBM to be at 1.51\,eV below $E_F$, also with no observation of the conduction band, and thus established 1.51\,eV as a lower bound (i.e. minimum possible value) of the electronic gap~\cite{BIANCHI2023BULK}. Yet, due to significant charging at low temperatures, the authors speculated that the gap is likely much larger than 1.51\,eV~\cite{BIANCHI2023BULK}. Our ARPES measurements are able to establish a larger lower bound on the electronic gap because our sample is heavily n-doped, as evidenced through comparison to our calculations. This eliminates (or significantly reduces) charging effects in our samples (See SI~\ref{fig:SM3}). This n-doping arises from our growth procedure where the high vapor pressure helps reduce halide vacancies.

Because our sample is heavily n-doped, we expect the conduction band to be just above $E_F$. Therefore, it is likely that the conduction band can be detected through \textit{in-situ} alkali metal surface doping (dosing), a procedure expected to donate electrons to the system, thus raising the Fermi level and causing a rigid downward energy shift of the bands. Using potassium (K) as our dopant, we performed five rounds of dosing to reach a surface dopant density of 0.35 electrons per unit cell (See SI Fig.~\ref{fig:SM2}). The successful dosing of K is confirmed through the growth of the K 3\textit{p} core peak at $\sim$\,19\,eV below $E_F$, as shown by the vertical dashed line in Fig.~\ref{fig:Fig2} (g). More importantly, a feature arises at $E_F$ upon dosing, highlighted by the black arrows in Fig.~\ref{fig:Fig2} (g). To explore this feature more closely, we present the high symmetry band dispersion along $\Gamma-$X and $\Gamma-$Y for fully dosed (0.35 electrons per unit cell) bulk CrSBr in Figs.~\ref{fig:Fig2} (c) and (d). In the dosed samples we observe a band near $E_F$ that is nearly flat along $k_x$ but highly dispersive along $k_y$. By comparing the dispersion to our sc\textit{GW} calculations, we conclude that this feature is the conduction band and not from other sources such as degenerate defect states. We note that the experimental band gap is indirect, as the conduction band minimum (CBM) is located at the X point, $\sim$50~meV lower than at $\Gamma$ (See SI Fig.\,\ref{fig:SM_Indirect}). Additionally, we observe that, besides broadening, the shapes of the valence bands remain qualitatively similar upon K dosing but that the energy of the top valence band has increased while the energies of the lower valence bands remain roughly unchanged. As the conduction band shifts downward while the valence band shifts upward, we necessarily observe a reduction in the band gap as a function of K dosing (Fig.~\ref{fig:Fig2} (h)). This evolution diverges from the expected rigid downward band movement and fixed band gap. 

\begin{figure} [ht]
    \includegraphics[width = 5 in]{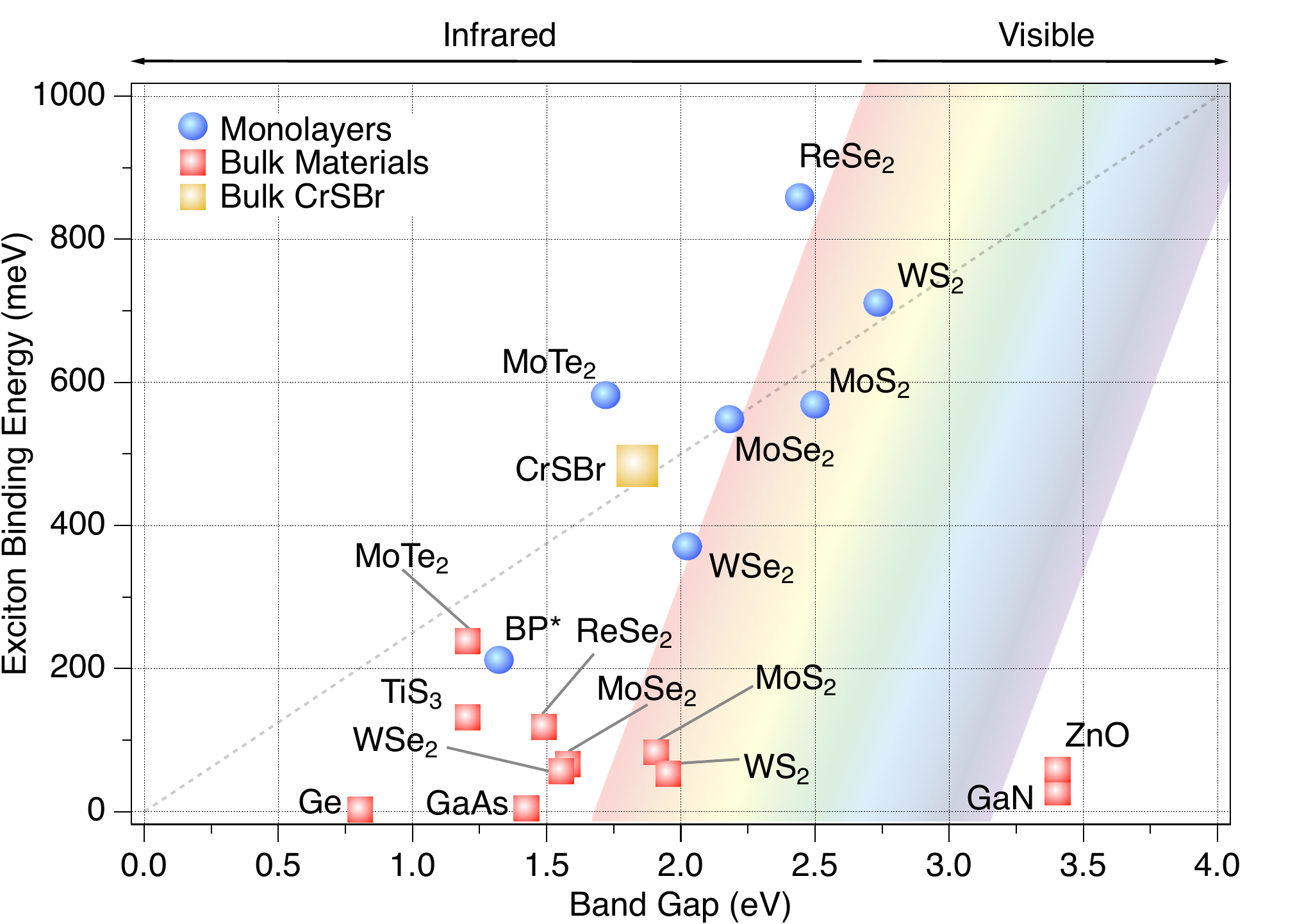}%
    \caption{\textbf{Exciton Binding Energies of Various Semiconductors.} 
    Plot of the exciton binding energy versus the electronic band gap for various monolayer and bulk semiconductors\,\cite{ZHANG2018,SAIGAL2016,BEAL1972,JUNG2022,BEAL1976,ARORA2015,NAM1976,DVORAK2013,MACFARLANE1959,ARORA2017_2,MOLINA2015,KLOTS2014,UGEDA2014,YANG2015,ZHU2014,HE2014,KANG2017}. 
    For the bulk TMDs, the gap at the K point (which is the relevant transition for the excitons) is plotted instead of the indirect gap. For all the other materials, the direct band gaps are used. For CrSBr, the gap plotted is the lower bound on the electronic gap found through our pristine ARPES measurements. The rainbow background represents the spectrum of the exciton energy (optical gap) for a given electronic gap and exciton binding energy. Notably, bulk materials have an exciton binding energy below 250~meV except for bulk CrSBr, which has a binding energy similar to those for monolayer TMDs. The gray dashed-line plots the $E_b\,=\,\frac{1}{4}E_{g}$ trend for monolayer semiconductors\,\cite{JIANG2017} which bulk CrSBr roughly obeys. All data points are from experimental works. The measurements for anisotropic black phosophorous (BP) were performed on a bilayer, hence the asterisk.
    \label{fig:Fig3}}
\end{figure}

Thus, our ARPES experiments have unveiled three notable findings: 1) a substantial band gap exceeding 1.84\,eV; 2) a highly anisotropic conduction band; and 3) an anomalous evolution in the band gap upon K dosing. We now explore the implications of each of these observations. 
First, it is evident that CrSBr hosts tightly bound excitons by considering previous optical studies that indicate an optical gap ($E_o$) of 1.32\,eV to 1.36\,eV for CrSBr at 200~K\,\cite{WILSON2021,WANG2023,LEE2021,LIN2024}. Comparing these values of $E_o$ to the lower bound of the electronic gap ($E_g$) extracted from our ARPES data, we establish the exciton binding energy ($E_b\,=\,E_g\,-\,E_o$) in single crystal CrSBr to be greater than 480\,meV. We also note that there is still debate surrounding the origin of the high-temperature optical signals. Thus, we compare our lower bound on $E_g$ below $T_N$ to the optical gap of 1.36\,eV found from our low-temperature reflectance measurements (SI~Fig.~\ref{fig:SM6} and SI~Fig.~\ref{fig:SM5}). From these low temperature measurements, we find an exciton binding energy exceeding 480\,meV, which is in agreement with our high temperature measurements. An exciton binding energy of this magnitude is significantly larger than other bulk inorganic semiconductors~\cite{ZHANG2018,SAIGAL2016,BEAL1972,JUNG2022,BEAL1976,ARORA2015,NAM1976,DVORAK2013,MACFARLANE1959,ARORA2017_2,MOLINA2015,KLOTS2014,UGEDA2014,YANG2015,ZHU2014,HE2014,KANG2017}, including bulk TMDs (Fig.~\ref{fig:Fig3}). Rather, this lower bound on the exciton binding energy roughly follows the universal trend of monolayer semiconductors where $E_b\,=\,\frac{1}{4}E_g$~\cite{JIANG2017}, suggestive of the strong van der Waals character of bulk CrSBr.

\begin{figure} [ht]
    \includegraphics[width = 6.3 in]{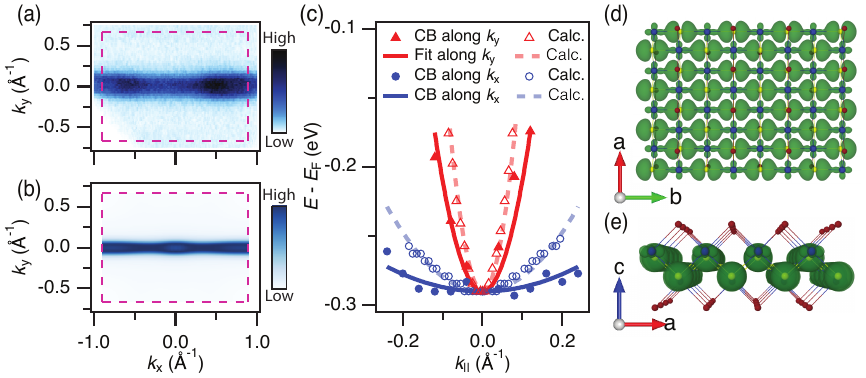}%
    \caption{\textbf{Quasi-1D Conduction Band in CrSBr} 
    (a) Experimental iso-energy plot in the $\Gamma$XSY plane at the conduction band energy at $\Gamma$ taken at $T\,=\,192$K (PM regime).
    (b) Calculated iso-energy plots in the $\Gamma$XSY plane at the conduction band minimum. 
    The stripe-like feature is indicative of a quasi-1D electronic structure.
    (c) Experimental (solid markers and lines) and calculated (empty markers and dashed lines) energy dispersion of the CB along $\Gamma-$X and $\Gamma-$Y lines. The markers are band peak positions found from peak positions of the EDCs. The lines are quadratic fits to the peak positions. The calculated curves have been offset in energy to match the position of the experimental data. 
    (d)-(e) sc\textit{GW} calculations of the spatial extent of the dominant orbitals of the bottom conduction band and top valence band viewed along the (d) \textit{c} and (e) \textit{b} axes. The electrons are localized in 1D chains along the \textit{b} axis.
    \label{fig:Fig4}}
\end{figure}

To explore the physical mechanisms for this exceptionally large bulk exciton binding energy, we examine the evolution of the electronic structure upon K dosing. First, we analyze the significant anisotropy of the conduction band to understand charge localization. Figure~\ref{fig:Fig4} (a) presents the iso-energy plot in the $\Gamma$XSY plane at an energy corresponding to the conduction band at $\Gamma$ (0.29\,eV below $E_F$). The presence of a single stripe-like feature that is extended along $k_x$ and truncated along $k_y$ confirms the in-plane anisotropic nature of the conduction band. This feature is in excellent agreement with our sc\textit{GW} calculation (Fig.~\ref{fig:Fig4} (b)). Additionally, we observe a stripe-like feature extended along $k_z$ and truncated along $k_y$, indicating a relatively flat conduction band along the $k_z$ direction (Fig.~\ref{fig:SM2} (e)). From these observations along with Figs.~\ref{fig:Fig2} (c) and (d), it is evident that the conduction band exhibits quasi-1D characteristics~\cite{CHEN2023_1D}. 

We quantify this observed anisotropy by fitting the peaks of the energy dispersion curves (EDCs) along $\Gamma-$X and $\Gamma-$Y to extract the effective mass of the electrons in the vicinity of $\Gamma$ (Fig.~\ref{fig:Fig4} (c)). We report experimental electron effective masses of $m_x^*\,=\,12.26\,m_e$ (3.58~$m_e$, calculated) and $m_y^*\,= \,0.48\,m_e$ (0.22~$m_e$, calculated), where $m_e$ is the free electron mass. These values correspond to an effective mass ratio of $\frac{m_x^*}{m_y^*} = 25.63$ (16.43, calculated). The discrepancy between experimental and calculated values is likely due to strong correlation effects outside the scope of our calculations. This effective mass anisotropy is large, even among other known quasi-1D materials~\cite{JIN2015,TRAN2014}. Notably, the mass anisotropy is significantly greater than previous reports of $\frac{m_x^*}{m_y^*} = 6.50$  in exfoliated CrSBr~\cite{BIANCHI2023THIN}, likely due to substrate effects in exfoliated CrSBr. Furthermore, calculations reveal significant anisotropy in the valence band as well with hole effective masses of $m_x^*\,=\,3.75\,m_e$ and $m_y^*\,= \,0.17\,m_e$ which correspond to a ratio of $\frac{m_x^*}{m_y^*} = 22.32$.  

This effective mass anisotropy in both the conduction and valence bands is indicative of localization of the charge carriers, which can motivate the large exciton binding energy. To confirm this charge localization, we calculated the real-space densities of the dominant orbitals comprising the lowest conduction and highest valence band (Figs. \ref{fig:Fig4}) (d) and (e)). We observe a strong charge localization along the crystallographic \textit{b} axis, consistent with previous calculations~\cite{KLEIN2023} and our observations in the electronic structure. We thus confirm robust confinement of charge carriers along 1D channels in bulk CrSBr, which contributes to the large observed exciton binding energy.

We next examine the anomalous evolution of the electronic band structure upon K dosing. From Figs.~\ref{fig:Fig2} (a)-(d), it is clear that in the K dosed dispersion, the conduction band has shifted downward in energy, while the valence band has shifted upward, relative to the pristine ones. By measuring the conduction and valence band positions as a function of K dosing, we track this evolution more closely (SI Fig.~\ref{fig:SM2} (d)). We observe that the conduction band suddenly appears after the first dosing round (0.07 electrons per unit cell). It then proceeds to shift only slightly downward in energy upon further dosing. In contrast, the valence band initially shifts downward before gradually moving upward. Importantly, we observe a significant reduction in the direct gap at $\Gamma$ from greater than 1.84~eV down to 1.65~eV at a dosing level of 0.07 electrons per unit cell before a roughly linear reduction down to 1.39~eV at 0.35 electrons per unit cell (Fig.~\ref{fig:Fig2} (h)). This tunability of the band gap with K dosing is consistent within multiple samples.

We now discuss the physical explanations for this anomalous behavior. We first rule out the possibility of negative electronic compressibility (NEC), a decreasing of the Fermi level upon an increase in carrier density, as we do not observe an upward energy shift in neither the deep valence bands nor the core levels~\cite{HE2015,WEN2023} (Figs.~\ref{fig:Fig2} (a)-(d), (g) and the traces for Sample 1 in SI Fig.~\ref{fig:SM3}). Additionally, the evolution cannot be explained by the intercalation of K atoms as the K 3\textit{p} core peak maintains its sharp line shape throughout the entire dosing process~\cite{ZHANG2019} (Fig.~\ref{fig:Fig2} (g)). Instead, we propose that the anomalous band evolution and reduction of the gap are due to carrier-induced band gap renormalization (BGR) and the surface Stark effect (SSE). 

First, band gap renormalization is a well-documented effect where the introduction of free carriers alters the screening environment and causes a sharp decrease in the electronic gap~\cite{FARIDI2021,GAO2017,LIU2019,LIANG2015}. Notably, this effect is theorized to be most pronounced for low doping levels and for a high density of states of the free carriers~\cite{GAO2017}. Indeed, our data demonstrate a large band gap reduction upon light K dosing and the quasi-1D nature of the conduction band supports a large electronic density of states. Furthermore, previous ARPES studies on low-dimensional CrSBr exfoliated on Au and Ag report band gaps of only 1.14 and 1.18eV~\cite{BIANCHI2023THIN}, respectively, which, we argue, is suggestive of substrate-induced BGR~\cite{SPATARU2010}. Additionally, a similar evolution of the valence and conduction bands has been observed in K-dosed WS\textsubscript{2}/h-BN where the increase in screening from the filling of the conduction band is found to be responsible for the band renormalization~\cite{KATOCH2018}. This is consistent with our BGR description of the band normalization we observe in CrSBr. Thus, we ascribe the sudden emergence of the conduction band and immediate decrease in gap size to carrier-induced BGR.

The linear upward shift in the valence band and resultant reduction in band gap for dosing levels greater than 0.07 electrons per unit cell can then also be explained by the SSE (Fig.~\ref{fig:Fig2} (h) and SI Fig.~\ref{fig:SM2} (d)). The SSE is a phenomenon whereby a vertical electric field localizes the conduction and valence electrons at different potentials in real space which results in a renormalization of the band gap~\cite{KHOO2004,RAMASUBRAMANIAM2011,ZHENG2008}. Notably, this effect has been throughly demonstrated in alkali metal-dosed van der Waals materials~\cite{ZHANG2019,KANG2017,KIM2015,FUKUTANI2019} where the ionized alkali metal atoms on the surface are responsible for the vertical electric field. Furthermore, a previous theoretical work has demonstrated that an external vertical electric field applied to CrSBr is capable of creating the strong real-space separation of the conduction and valence electrons necessary to produce the SSE~\cite{DANG2022}. Additionally, the band gap reduction from the SSE is expected to be linear as a function of electric field for large fields~\cite{KHOO2004,RAMASUBRAMANIAM2011,ZHENG2008}, in agreement with our ARPES data. We also note that competition between the chemical potential shift and the SSE has been previously reported and can explain the initial downward movement of the valence band~\cite{KIM2015}.

Importantly, large contributions from both carrier-induced BGR and SSE are indicative of weak dielectric screening and strong 2D character~\cite{LIU2017, FARIDI2021, GAO2017,KANG2017,ISHIGAMI2005,KATOCH2018}, factors which encourage strong Coulomb interactions and enhance the exciton binding energy. It is also essential to recognize that while the SSE separates the valence and conduction states in real space, the structural confinement ensures that excitons can remain tightly bound\,\cite{MILLER1984,WALTERS2018} and thus such an effect is consistent with the pronounced excitonic behavior in CrSBr.

We therefore determine that bulk CrSBr hosts tightly bound excitons due to quasi-1D charge localization and weak dielectric screening. The key role of the anisotropy, both structural and electronic, provides optimism that large exciton binding energies can be found in other highly anisotropic bulk van der Waals systems. Particularly, other van der Waals magnets, where predictions of extraordinarily large exciton binding energies in CrBr\textsubscript3, CrI\textsubscript3, and MnPS\textsubscript3 have already been made~\cite{BIROWSKA2021,GRZESZCZYK2023}, could provide additional systems in which to study the dimensionality of excitons. The band gap tunability under an applied electric field also presents a method to easily adjust the electronic and optical properties of CrSBr\,\cite{MILLER1984,WALTERS2018,ZHOU2022}, an effect that opens possibilities for both the further study of many-body physics and the development of semiconductor devices.

\newpage

\section{Data availability}
Data underlying these results are available from the authors, see Additional Information below. 

\section{References and notes}

\normalem
%

\section{Acknowledgement}

We are grateful to Kai Sun for insightful discussions. This research was supported by the National Science Foundation (NSF) through the Materials Research Science and Engineering Center at the University of Michigan, Award No. DMR-2309029. This material is also based upon work supported by the NSF CAREER grant under Award No. DMR-2337535. This work used resources of the Advanced Light Source, a U.S. Department of Energy (DOE) Office of Science User Facility under Contract No. DE-AC02-05CH11231. Q.L., A.A and H.D. acknowledge the support by the Army Research Office Award No. W911NF-17-1-0312 and Gordon and Betty Moore Foundation Award No. GBMF10694. L.Z. acknowledges support by US Air Force Office of Scientific Research (AFOSR) YIP grant no. FA9550-21-1-0065, NSF CAREER grant no. DMR-174774 and Alfred P. Sloan Foundation. The work at University of Texas at Dallas is supported by the US AFOSR (FA9550-19-1-0037), NSF (DMREF-2324033) and Office of Naval Research (ONR) (N00014-23-1-2020).

\section{Author contributions}
W.L, A.L.N.K, and B.L. synthesized the single crystal. S.S., E.D., A.B., C.J., E. R., and N.H.J conducted the ARPES experiment. M.W. performed the scGW calculations.  D.Z. and E.G. developed the scGW code and methodology.
Q.L., A.A., L.Z., and H.D. helped prepare samples and performed optical measurements. S.S., M.W., Q.L., D.Z., E.G., and N.H.J. wrote the manuscript in consultation with E.D., A.A., W.L., A.L.N.K., A.B., C.J., E.R., L.Z., and H.D.

\section{Competing financial interests}
The Authors declare no Competing Financial or Non-Financial Interests.

\section{Methods}
\subsection{Crystal Growth.} 

The CrSBr crystals used in this study were prepared in the exact manner described in Liu et. al.\cite{LIU2022}.

\subsection{ARPES Measurements}

All ARPES measurements were performed at Beamline 7.0.2 (MAESTRO) of the Advanced Light Source. The beamline is equipped with a R4000 spectrometer with deflectors that enable data collection across the full Brillouin zone without moving the sample. Bulk CrSBr crystals were mounted on Cu pucks with Epotek H20E silver epoxy and cleaved \textit{in situ} at vacuum better than $5\times10^{-11}$ mbar. All measurements with fixed photon energy were performed with 83-84eV photons with linear horizontal polarization. The beam spot size was 15 $\mu$m x 15 $\mu$m. Paramagnetic measurements were performed at temperatures of 192K, 195K, and 195K for Sample 1, Sample 2, and Sample 3, respectively while antiferromagnetic measurements (SI Fig.~\ref{fig:SM5}) were performed on Sample 1 at a temperature of 97K. Potassium dosing experiments were performed by evaporating potassium onto the cleaved CrSBr surface $in situ$ from a SAES getter source such that the sample was not moved from the measurement position. The level of potassium adsorption was estimated using Luttinger’s theorem (see SI for full details). 

\subsection{Self-consistent \textit{GW} calculations}
Self-consistent GW \cite{Hedin65} calculations were performed on the Matsubara axis using a finite-temperature Gaussian-orbital based self-consistent Green's function solver \cite{Iskakov20,yehFullySelfconsistentFinitetemperature2022a,Green23}. Calculations used the gth-szv-molopt-sr basis set \cite{VandeVondele07} with the gth-pbe pseudopotential \cite{Goedecker96}.

Self-consistent $GW$ iterations were initialized with a density functional calculation using the PBE functional \cite{Perdew96}, using the \texttt{pyscf}~\cite{sunPySCFPythonbasedSimulations2018,sunRecentDevelopmentsPySCF2020} software package, and iterated to self-consistency.

Results were converged on the Matsubara axis in a sparse sampling \cite{Li20} intermediate representation  \cite{Shinaoka17} grid using convergence acceleration  \cite{Pokhilko22}. Results were then analytically  continued to the real frequency axis using Nevanlinna analytical continuation.~\cite{feiNevanlinnaAnalyticalContinuation2021}.

The results shown here are obtained in a monolayer geometry using a periodic $k$-space mesh of size $6\times8\times1$. 
Supporting Materials present additional details regarding the symmetrized atomic orbital decomposition, charge density calculation, and 2-D density of state iso-energy surface.

\newpage

\section{Supplementary Information}

\renewcommand{\thepage}{S\arabic{page}}  
\renewcommand{\thesection}{S\arabic{section}}   
\renewcommand{\thetable}{S\arabic{table}}   
\renewcommand{\thefigure}{S\arabic{figure}}
\setcounter{figure}{0}
\setcounter{table}{0}

\subsection{Determination of high symmetry points}

\begin{figure} [ht]
    \includegraphics[width = 6.3 in]{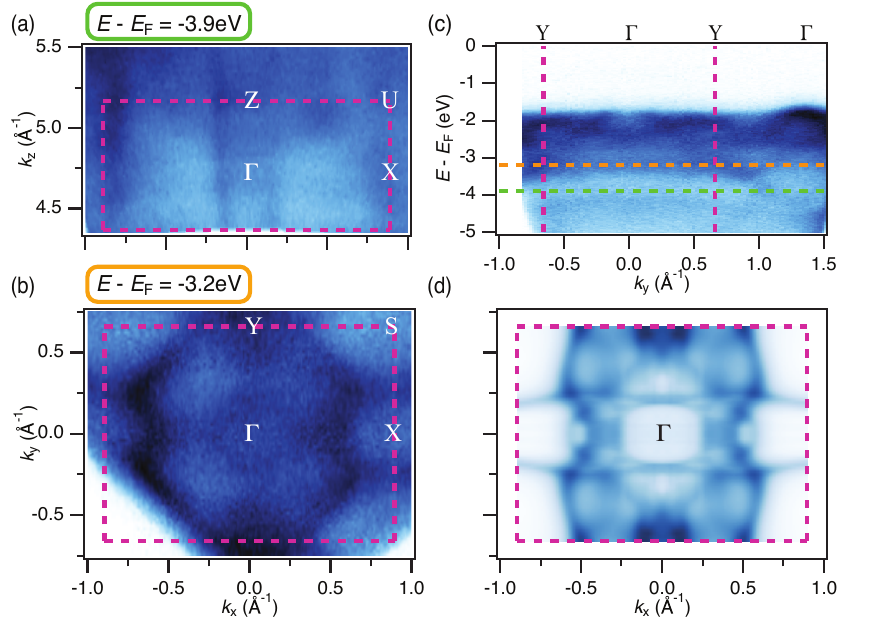}%
    \caption{\textbf{Valence Band Dispersion of Pristine CrSBr.} 
    (a) Out-of-plane iso-energy plot taken at a binding energy of 3.9\,eV. A stripe-like feature is observed along $k_z$ around $\Gamma$ demonstrating the weak out-of-plane dispersion expected for a van der Waals material. However, slight curvature to this stripe feature demonstrate the periodicity. (b) In-plane iso-energy plot taken at a binding energy of 3.2\,eV. The expected two-fold symmetry is observed and the periodicity aligns with the expected lattice constants. (c) Band dispersion plot along the $\Gamma-$Y high-symmetry line. The VBM is visible in the second Brillouin zone. The green and orange dashed lines denote the energies at which (a) and (b) are respectively taken. (d) sc\textit{GW} calculation of the iso-energy plots shown in (b). The pink dashed lines represent the first Brillouin zone boundaries.
    \label{fig:SM1}}
\end{figure} 

\begin{figure} [ht]
    \includegraphics[width = 6.3 in]{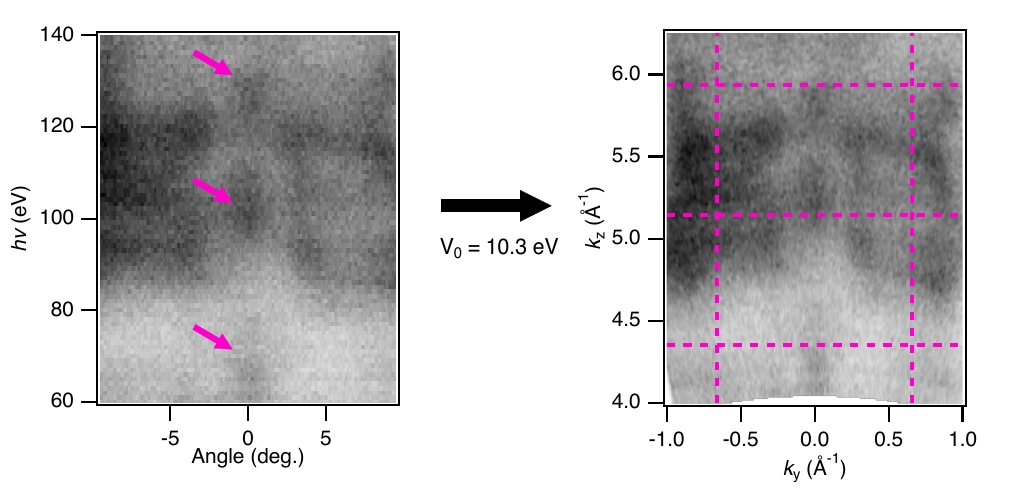}%
    \caption{\textbf{Brillouin Zone Orientation.} An illustration of the determination of the out-of-plane high symmetry points in bulk CrSBr. The left image is a photon energy dependent measurement in the $k_x\,=\,0\,\text{\AA}^{-1}$ plane. The periodic intensity pulses along the 0 degree line, highlighted by the pink arrows, are used to determine the Brillouin zone orientation. Using the nearly-free electron approximation and the observed symmetry, the photon energies are converted to $k_z$ values with an inner potential of 10.3\,eV as shown on the right. Both plots are taken at a binding energy of 3.65\,eV. The pink dashed lines represent the Brillouin zone boundaries under this transformation.
    \label{fig:SM_BZ}}
\end{figure} 

We explore the electronic structure of the valence bands by examining the in- and out-of-plane dispersion throughout the Brillouin zone. To probe the out-of-plane ($k_z$) dispersion, we performed photon energy dependent ARPES measurements. We observe minimal dispersive features in the out-of-plane measurements indicative of the quasi-2D electronic structure expected for a van der Waals material. However, by focusing near the top of the band peaking at roughly 4\,eV below $E_F$, we are able to distinguish a slight periodic feature in the out-of-plane iso-energy plot (SI~Fig.~\ref{fig:SM1} (a)). This feature centered at $k_x\,=\,0$ is still stripe-like along the $k_z$ direction but has slight curvature that makes a repeating pattern visible. Using this periodic pattern, as well as a more pronounced periodic pattern observed in a perpendicular cut (Fig.~\ref{fig:SM_BZ}), we identify the center of the Brillouin zone to be at a photon energy of 83-84\,eV. We calculate the corresponding $k_z$ values using the final state nearly-free electron approximation with an inner potential of $V_0\,=\,10.3\,eV$ (Fig.~\ref{fig:SM_BZ}). This compares to the previous ARPES study in which a photon energy of 79-80\,eV corresponded to the center of the Brillouin zone with an inner potential of 12.8\,eV~\cite{BIANCHI2023BULK}. Comparison between their determination of the symmetry points in $k_z$ and ours reveals a discrepancy of roughly $0.13\,\text{\AA}^{-1}$ in $k_z$. Due to a lack of out-of-plane dispersion arising from the planar crystal structure, the band structure is relatively unchanged between photon energies of 80\,eV and 87\,eV (a range of $\sim0.20\,\text{\AA}^{-1}$ in $k_z$) with energy variations in the lowest conduction band and top valence band of only $\sim30$\,meV, suggesting that such a variation has minimal effect on our reports of the electronic structure.

We locate the in-plane high symmetry points through an iso-energy plot in the $\Gamma$XSY plane using a photon energy of 83eV (Fig.~\ref{fig:SM1} (b)). We clearly observe the two-fold symmetry expected for the orthorhombic unit cell. Importantly, the periodicity aligns  with the in-plane expected lattice constants. Our in-plane experimental data are in good agreement with our sc\textit{GW} calculations (Fig.~\ref{fig:SM1} (d)), further suggesting we have properly aligned our sample and found the high-symmetry points. Additionally, we note that the top valence band energy position only varies by $\sim40$\,meV in a range of $\pm\,0.1\,\text{\AA}^{-1}$ in $k_y$ around $\Gamma$.

\subsection{Measurement of valence band maximum} 

To quantify the lower bound of the band gap of pristine bulk CrSBr, we measure the binding energy of the valence band maximum (VBM). As discussed in the main text, our calculations predict the VBM to be at $\Gamma$. While we do not observe the VBM in the first Brillouin zone due to matrix element effects, we do indeed find the VBM at $\Gamma$ in the second Brillouin zone along $k_y$ (Fig.~\ref{fig:SM1} (c)).

\newpage

\subsection{Quantification of K dosing}

We estimate the amount of additional electron concentration through the size of the electron pockets at the Fermi level following the procedure in Zhang et. al.\cite{ZHANG2021}. To do so, we use Luttinger’s theorem\cite{Luttinger1960}:
					\[\Theta\,=\,2\,\times\,\frac{A_{EP}}{A_K}\]
where $\Theta$ is the electron density in electrons per unit cell, $A_{EP}$ is the area of the electron pocket at the Fermi surface, $A_K$ is the area of the surface Brillouin zone, and the factor of 2 is from the electron degeneracy. We approximate the electron pockets as rectangles that span the entire Brillouin zone along $k_x$ with a finite width in $k_y$ that we determine from the momentum distribution curves (MDCs) at the Fermi surface.
Using this procedure, we estimate the \textit{k}-space area of the electron pockets for fully-dosed Samples 1, 2, and 3 to be 0.412, 0.125, and 0.435 $\text{\AA}^{-2}$, respectively. This corresponds to $\Theta$ of 0.35, 0.11, and 0.37 electrons per unit cell for our respective samples. Dividing by the area per unit cell of $4.75\,\times\,3.50\,=\,16.625\,\text{\AA}^{2}$ results in electron densities $n$ of $2.09\,\times\,10^{14}$, $6.37\,\times\,10^{14}$, and $2.21\,\times\,10^{14}$ electrons/cm$^2$. To convert this electron density to K coverage, we assume that the adsorbed K forms a lattice with unit cell area off $15.205\,\text{\AA}^{2}$ (from K radius of $2.2\,\text{\AA}$)\,\cite{ZHANG2021}. Thus, one monolayer corresponds to an electron density of $6.58\,\times\,10^{14}$ electrons per square cm. By dividing the estimated sample electron density by this value, we therefore estimate the maximum achieved K coverage for each of Sample 1, 2 and 3 to be 0.32, 0.10, 0.34\,ML, respectively. We then estimate the magnitude of the intermediate dosing levels by interpolating given the number of dosing rounds and assuming a fixed deposition rate for a given set of input parameters for the potassium getter.

\subsection{Dosing evolution}

\begin{figure} [ht]
    \includegraphics[width = 6.3 in]{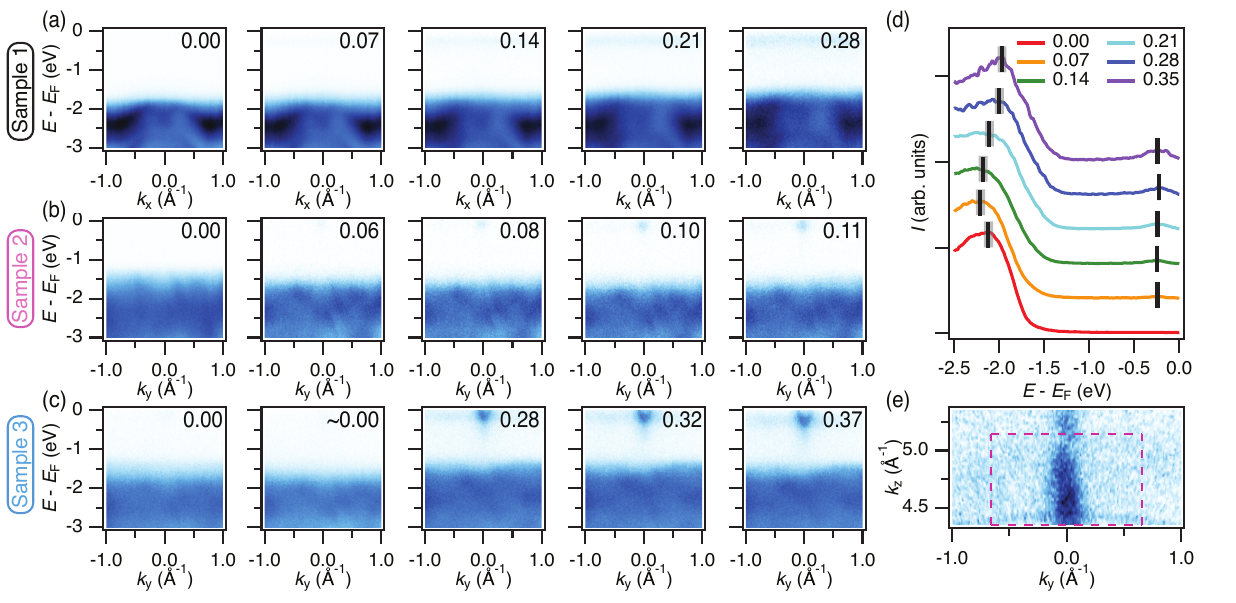}%
    \caption{\textbf{Full K Dosing Evolution of the Electronic Structure of CrSBr} 
    (a-c) Evolution of the energy dispersion of CrSBr as a function of K dose for (a) Sample 1, (b) Sample 2, and (c) Sample 3. The energy dispersion plots in (a) are taken along the non-symmetric line $k_y\,=\,0.29\,\text{\AA}^{-1}$ while (b,c) are taken along the $\Gamma-$Y high symmetry line. In all samples, the conduction band emerges upon K dosing. (d) EDCs at $k_x\,=\,0\,\text{\AA}^{-1}$, $k_y\,=\,0.29\,\text{\AA}^{-1}$ at various levels of K dose for Sample 1. (e) Experimental iso-energy plot in the $\Gamma$YTZ plane at the conduction band (CB) minimum for Sample 2. The stripe-like feature confirms the lack of dispersion along $k_z$.
    \label{fig:SM2}}
\end{figure}

We performed K dosing ARPES studies on each of our three samples to reveal the conduction band and explore its evolution (Methods). We applied K to our samples incrementally and measured the core spectrum (
Fig.~\ref{fig:Fig2} (g)) and band dispersion after each dose. A detailed evolution of the electronic structure as a function of K dose for all three samples is shown in Figs.~\ref{fig:SM2} (a)~-~(c). The band dispersions are taken along the $\Gamma-$Y high symmetry cut for Samples 2 and 3 while the dispersion for Sample 1 is taken along a non-symmetric line along $k_x$ at $k_y = 0.29\,\text{\AA}^{-1}$. The dispersion for Sample 1 is taken along this non-symmetric line because we did not properly align the detector prior to the measurement. In all three samples, the conduction band emerges and then shifts downward in energy as K is applied. The valence band, however, undergoes an anomalous shift, evident through visual comparison in Figs.~\ref{fig:SM2} (a)~-~(c). 

To quantify the movement of the valence and conduction bands, we tracked their peak positions for each of the three samples using energy dispersion curves (EDCs) taken after each dosing round. A representative analysis of the EDCs for Sample 1, taken at $k_x\,=\,0\,\text{\AA}^{-1}$, $k_y\,=\,0.29\,\text{\AA}^{-1}$, is shown in Fig.~\ref{fig:SM2} (d). For pristine bulk CrSBr, no peak is observed for the conduction band. Upon the first dosing round (0.07 electrons per unit cell), the valence band peak clearly shifts downward while a peak arises near the Fermi level. Upon further dosing (0.07 to 0.35 electrons per unit cell), the top valence band gradually shifts upward in energy while the conduction band does not shift noticeably. Due to the minimal in- and out-of-plane dispersion around $\Gamma$, as detailed in the previous section, such band evolution is not due to a misalignment in the Brillouin zone. Importantly, we confirm that the conduction band is quasi-1D by examining its out-of-plane dispersion (Fig.~\ref{fig:SM2} (e)). The stripe-like feature extended along $k_z$ and truncated along $k_y$ indicates that the conduction band is flat along $k_z$ (as well as $k_x$, as discussed in the main text), confirming the quasi-1D nature.

We calculated the gap at $\Gamma$ (Shown in Main Text Fig.~\ref{fig:Fig2} (h)) by finding the difference between the energies of the conduction band and valence band positions at $\Gamma$ as found through the EDC analysis described above. 
While our EDCs for Sample 1 were not taken at $\Gamma$, we estimated the gap through a careful transformation. First, we assumed that the difference between the valence band at $\Gamma$ and the valence band at $k_x\,=\,0\,\text{\AA}^{-1}$, $k_y\,=\,0.29\,\text{\AA}^{-1}$ is fixed (i.e. independent of dosing). This assumption is justified by the fact that the difference between valence band position at $\Gamma$ and at $k_x\,=\,0\,\text{\AA}^{-1}$, $k_y\,=\,0.29\,\text{\AA}^{-1}$ is unchanged within the margin of error between pristine (a difference of 0.28\,eV) and fully dosed (0.29\,eV) measurements. Next, while we cannot observe the conduction band at $k_x\,=\,0\,\text{\AA}^{-1}$, $k_y\,=\,0.29\,\text{\AA}^{-1}$, we do observe diffuse scattering from these states caused by the disordered surface induced by the K adsorption. Such scattering forms a flat feature throughout the Brillouin zone at an energy around the conduction band minimum. This was observed in all three samples. In Sample 1, this flat feature from scattering appears 0.05\,eV above the conduction band minimum. Therefore, we assume that the energy position of this scattered feature is a fair estimate for the energy of the conduction band at $\Gamma$. Thus, the gap at $\Gamma$ for Sample 1 was estimated by subtracting 0.34\,eV from the gap measured at $k_x\,=\,0\,\text{\AA}^{-1}$, $k_y\,=\,0.29\,\text{\AA}^{-1}$. For Samples 2 and 3, the EDCs were taken at $\Gamma$ and thus the reported gaps at $\Gamma$ were measured directly. Larger error bars are added to the gap values for Sample 1 to take the transformation uncertainty into account.

\subsection{Determination of the indirect gap}

\begin{figure} [ht]
    \includegraphics[width = 6.3 in]{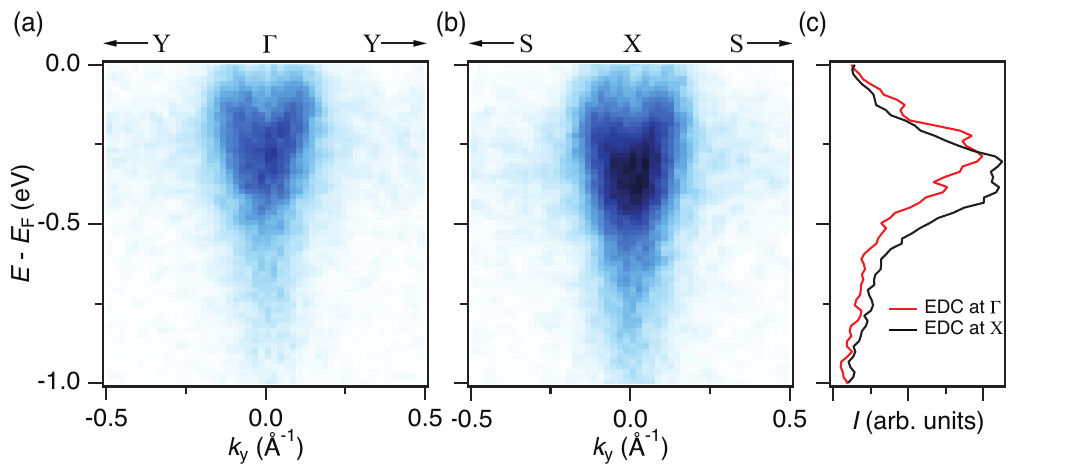}%
    \caption{\textbf{Indirect Bandgap of Bulk CrSBr} 
    (a-b) Conduction band dispersion along $k_y$ around the high symmetry points (a) $\Gamma$ and (b) X. (c) Energy dispersion curves at $\Gamma$ and X in an energy range focused at the conduction band. The conduction band is evidently lower in energy at X.
    \label{fig:SM_Indirect}}
\end{figure}

By comparing the gap at $\Gamma$ to the gap at other points in the Brillouin zone, we determine that bulk CrSBr has an indirect bandgap. Figures~\ref{fig:SM_Indirect} (a) and (b) depict the conduction band dispersion along $k_y$ around $\Gamma$ and X, respectively. The parabolic band is visibly lower in energy at X than at $\Gamma$. This is confirmed through analysis of the EDCs, where the nadir of the conduction band is at -0.35\,eV at X and at -0.29\,eV at $\Gamma$ (Fig.~\ref{fig:SM_Indirect} (c)). Since the valence band maximum is at $\Gamma$, this establishes bulk CrSBr as an indirect bandgap semiconductor.

\subsection{Cr 3p core peaks before and after dosing}

\begin{figure} [ht]
    \includegraphics[width = 6.3 in]{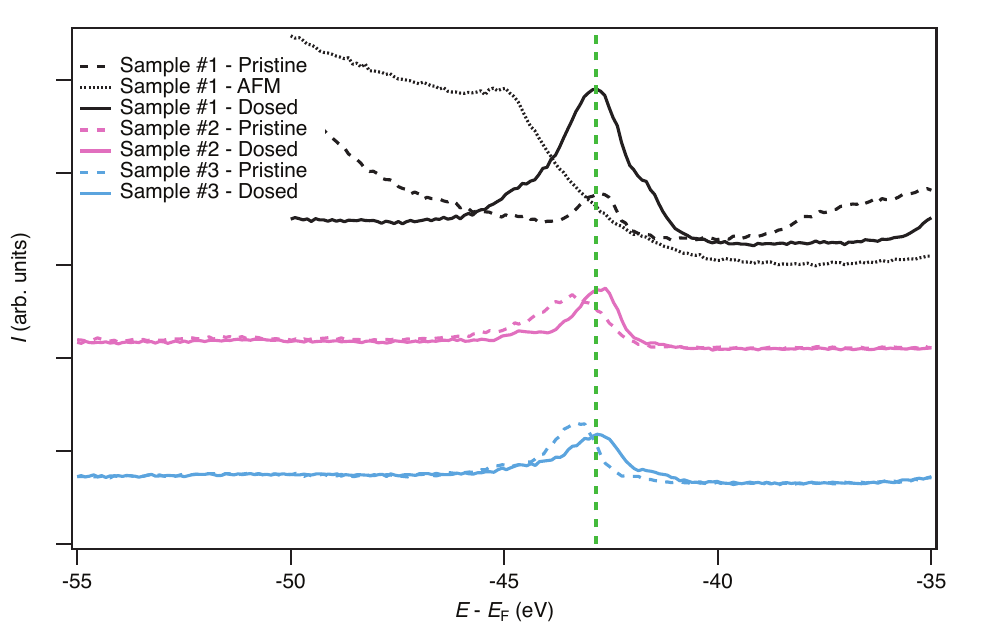}%
    \caption{\textbf{Cr 3\textit{p} Core Level Analysis} 
    X-ray photoemission spectroscopy (XPS) measurements of the Cr 3\textit{p} core level for pristine, K dosed, and AFM (Sample 1 only) bulk CrSBr for each of our three samples.
    \label{fig:SM3}}
\end{figure}

Due to the insulating nature of CrSBr, we observed varying levels of electrostatic charging during our ARPES measurements. Figure~\ref{fig:SM3} displays the Cr 3\textit{p} core peaks before and after dosing for each of the three samples. The pristine Cr 3p positions for Samples 1, 2, and 3 are $-42.87\,\pm\,0.1$, $-43.45\,\pm\,0.1$, and $-43.34\,\pm\,0.1$eV, respectively. The fully dosed positions are $-42.88\,\pm\,0.1$, $-42.80\,\pm\,0.1$, and $-42.84\,\pm\,0.1$eV, respectively. Because dosing renders the samples metallic, the core peak positions for dosed samples are taken as the true, uncharged positions. Thus, we conclude that the uncharged Cr 3\textit{p} core peak is roughly 42.85\,eV below $E_F$. This demonstrates that Sample 1 did not experience significant charging in its pristine state whereas Samples 2 and 3 did. To correct for the charging and find the band positions relative to $E_F$, we rigidly shifted the measured band positions of Samples 2 and 3 up in by an energy equal to the difference between the measured Cr 3\textit{p} peak positions and -42.85\,eV. This correction procedure adds additional uncertainty to the band positions of Samples 2 and 3. The values of the electronic gap for Samples 2 and 3 (Fig.\,\ref{fig:Fig2} (h)), however, are found from the difference between the valence and conduction band and thus these values do not suffer from any uncertainty induced by the charge correction. Importantly, Sample 1, which did not experience such charging (and from where the majority of our presented data derives), did not require this additional processing.

\subsection{Temperature dependence}

\begin{figure} [ht]
    \includegraphics[width = 6.3 in]{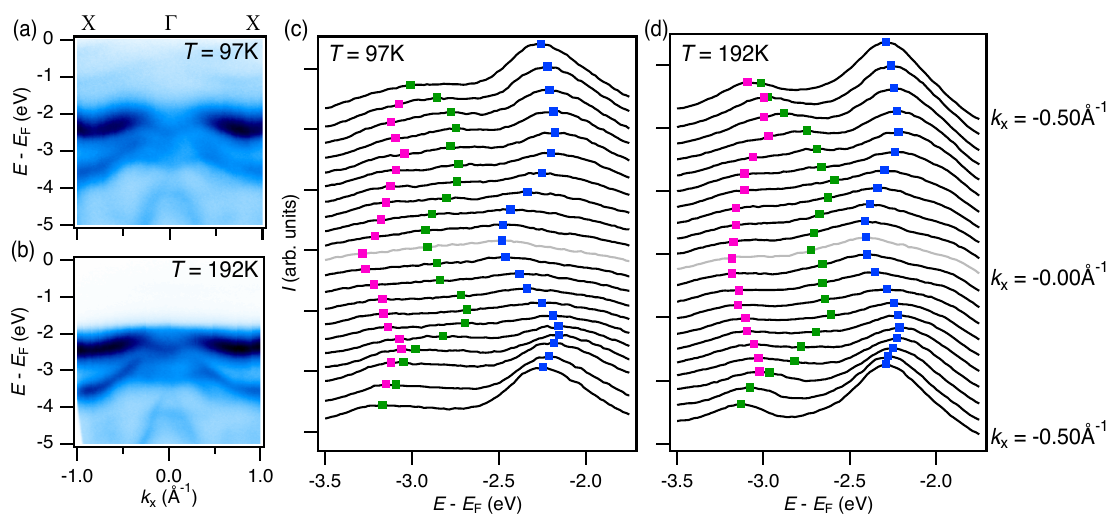}%
    \caption{\textbf{Temperature Dependence of the Electronic Structure of Bulk CrSBr} 
    (a,b) Energy dispersion plots taken along the $\Gamma-$X high symmetry lines for bulk CrSBr in the (a) AFM and (b) PM phases. (c,d) EDCs along the $\Gamma-$X high symmetry line from $k_x\,=\,-0.50\,\text{\AA}^{-1}$ to $k_x\,=\,0.50\,\text{\AA}^{-1}$ in steps of $0.05\,\text{\AA}^{-1}$. The peak positions of the top three valence bands, found through multipeak fitting of Lorentzians, are denoted by the markers. Notably, the trough at $\Gamma$ of the top two  bands is exaggerated and the gap between the second and third bands is diminished in the AFM phase.
    \label{fig:SM5}}
\end{figure}

To explore the effects of the magnetic ordering on the electronic structure of bulk CrSBr, we performed ARPES measurements at $T\,=\,97\,K\,<\,T_N$ (Fig.~\ref{fig:SM5}). Notably, we observed significant charging effects that resulted in a rigid downward shift of the bands. To correct for the charge-induced shift, we compared the Cr 3\textit{p} core peak in the AFM regime to that in the paramagnetic (PM) regime (Fig.~\ref{fig:SM3}). We found a difference of 2.13\,eV and thus applied this energy offset to correct the AFM spectra. Having done this correction, we then compare the electronic structure along the high symmetry $\Gamma-$X cut for AFM and PM bulk CrSBr (Figs.~\ref{fig:SM5} (a) and (b)). Note that the non-zero intensity visible in Fig.~\ref{fig:SM5} (a) is due to inhomogeneous charging and thus does not represent additional bands in the AFM regime. Qualitatively, much of the band structure along this cut is unchanged between the two phases. However, small differences are apparent in the top three valence bands around $\Gamma$. To analyze these differences more closely, we plot EDCs around $\Gamma$ for both the AFM (Fig.~\ref{fig:SM5} (c)) and PM (Fig.~\ref{fig:SM5} (d)) phases. Using a multi-peak fitting analysis, we plot the peak positions for each of the top three valence bands around $\Gamma$. Importantly, we observe that the trough around $\Gamma$ is exaggerated for the top two bands (blue and green markers in Figs.~\ref{fig:SM5} (c) and (d)) in the AFM phase. Specifically, the energy difference from the top of the upper valence band (blue markers) to the minimum at $\Gamma$ is 320\,meV in the AFM phase and 190\,meV in the PM phase. The third valence band (pink markers), however, is unchanged upon the magnetic transition. Thus, the gap between the second and third bands is reduced from a maximum of 520\,meV in the PM phase to a maximum of 380\,meV in the AFM phase. We also note that the local maximum of the valence band on these dispersion plots only differs by less than $\sim60$\,meV with the AFM dispersion being slightly higher in energy. Since we also observe significant charging in the AFM phase, this also implies that the conduction band has shifted upwards in energy. Since the valence band positions only changes slightly while significant charging suggests a larger (upward) shift of the conduction band, we argue that the electronic gap is not reduced in the AFM phase compared with the PM phase.
\newpage

\subsection{Optical measurements}

\begin{figure} [ht]
    \includegraphics[width = 3.5in]{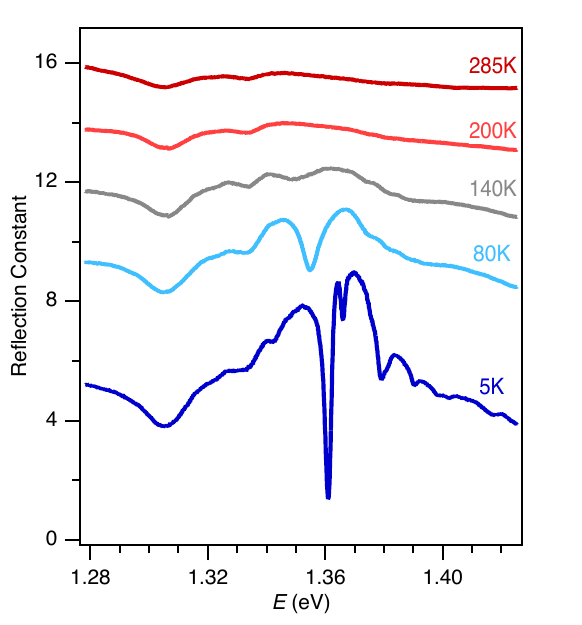}%
    \caption{\textbf{Temperature-Dependent Reflection Contrast Spectra of Bulk CrSBr} 
    Plot of the reflection contrast (RC) of bulk CrSBr as a function of photon energy for various temperatures between 5 and 285\,K. At 5\,K, the exciton energy is observed in the sharp trough at $E\,=\,1.361$\,eV. This feature redshifts, broadens, and decreases in amplitude with increasing temperature, disappearing above 140\,K.
    \label{fig:SM6}}
\end{figure}

To measure the optical gap, we performed reflection contrast (RC) measurements on bulk CrSBr samples on a SiO\textsubscript{2}/Si substrate. Figure~\ref{fig:SM6} presents the RC as a function of incident photon energy where the RC is defined as \[RC\,=\,\frac{R_{CrSBr}}{R_{SiO_2}}\,-\,1,\] with $R_{CrSBr}$ ($R_{SiO_2}$) being the reflection of CrSBr (SiO\textsubscript{2}). From these data, we observe an exciton energy of 1.361\,eV at 5\,K. As the temperature is increased, the exciton energy undergoes a redshift to roughly 1.349\,eV at 140\,K. Additionally, the exciton trough broadens and decreases in amplitude with increasing temperature, ultimately disappearing above 140\,K. 
While we are not able to observe the optical gap above $T_N$ using RC, photoluminescence (PL) does show a peak up to room temperature~\cite{WILSON2021,WANG2023}. At 200\,K, these measurements find the optical gap to be $\sim1.32-1.36$\,eV for bulk CrSBr~\cite{WILSON2021}. 
However, it is important to note that reflectance, which depends on oscillator strength, is a more reliable measure of the exciton energy than PL, which can have below-gap contributions from factors such as defect states or charged excitons. Thus, while excitons may be present above $T_N$, our RC measurements reveal with certainty the presence of excitons below $T_N$.
In either case, we find the exciton binding energy to exceed 480\,meV in single-crystal CrSBr. We also note that, in general, one has to be careful in comparing ARPES data with optical data because the final states are different in the two techniques (charge neutral in optical data versus a charged hole in ARPES). This can renormalize the band gap, but such an effect is typically small and thus has a minor impact on determination of the exciton binding energy.

\subsection{Self-Consistent \textit{GW} Calculations}

\textit{Density-of-state iso-energy surfaces}
Two-dimensional densities of state were obtained from the converged GW runs in the following way. First, Matsubara Green's functions were orthogonalized in symmetrized atomic orbitals at each momentum point calculated. Second, using a Wannier interpolation, Matsubara Green's functions were obtained on a fine two-dimensional momentum mesh. Nevanlinna analytical continuation was then used to analytically continue the Green's functions to real frequency and evaluate them at the desired energy.

For the $\Gamma$YSX plane, a $33 \times 45$ mesh in reciprocal space spanning a rectangle between the origin and the nearest Y and X points was used and energies were scanned between -0.7Ha and 0.3Ha.

\begin{figure}
    \centering
    \includegraphics[width = 1.0\linewidth]{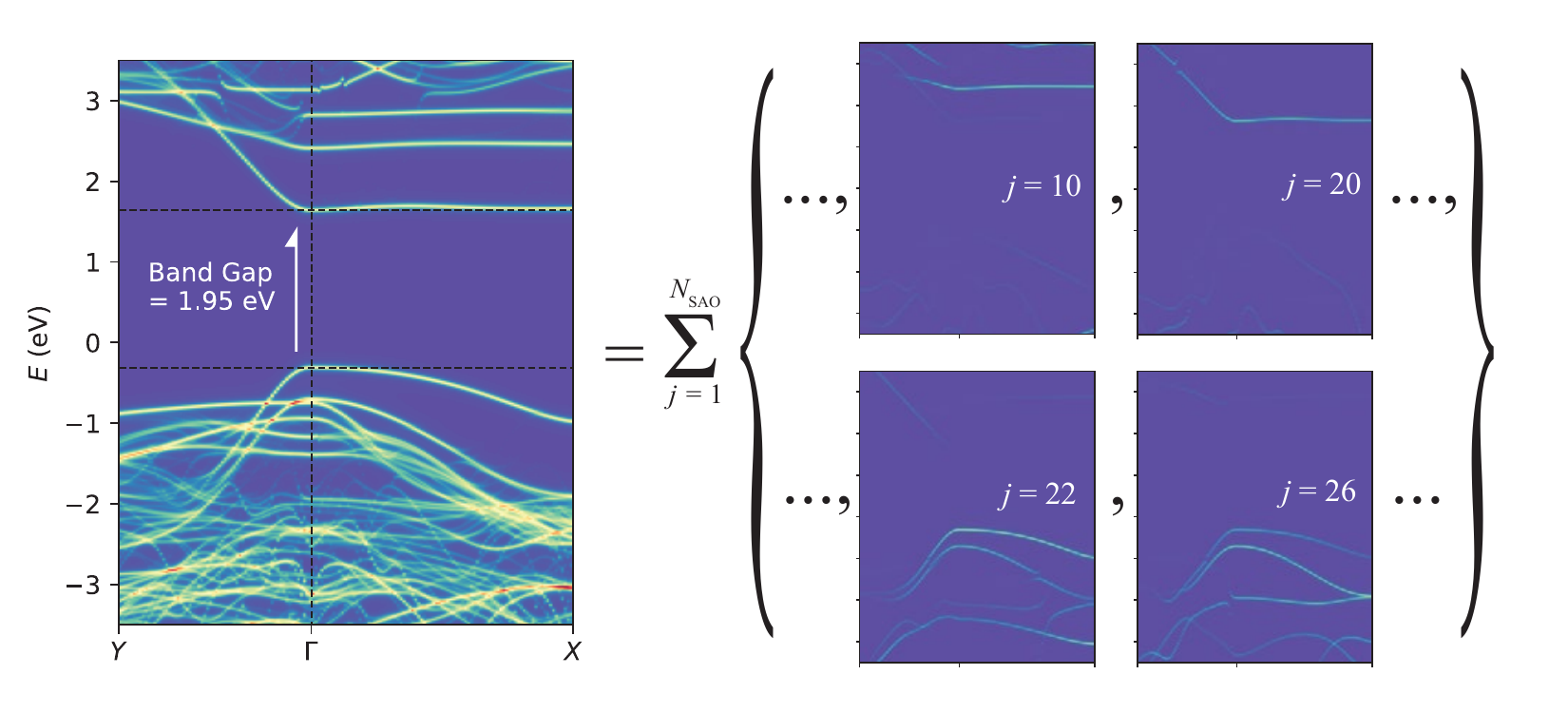}
    \caption{The monolayer CrSBr band structure decomposed as the sum of all SAOs. The four components plotted on the right ($j = 10,20,22,26$) are the most important SAOs that contributed to the first two virtual bands valence bands near the Fermi level at $\Gamma$.}
    \label{fig:SAO decomp bands}
\end{figure}

\begin{figure}
    \centering
    \includegraphics[width = 0.9\linewidth]{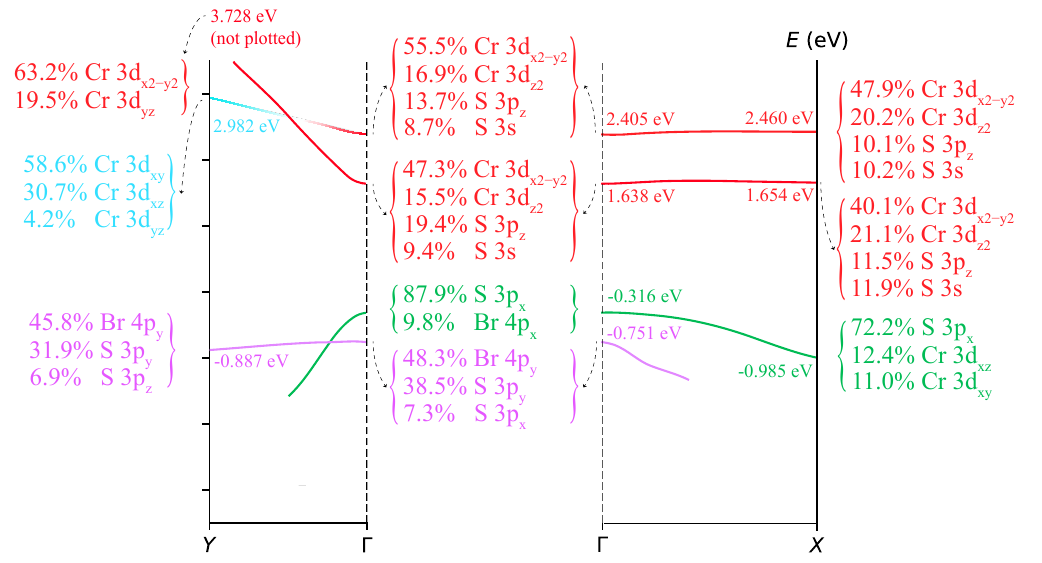}
    \caption{AO contribution percentages to the high symmetry points along the selected four bands near the Fermi level for the monolayer CrSBr. Minor AO contributions are omitted. The colors represent different dominant AOs. }
    \label{fig:AO decomp bands}
\end{figure}

\textit{Orbital decomposition into Symmetric Atomic Orbitals (SAO)} Calculations were performed in a set of Gaussian orbitals. Upon convergence of the calculation, the band structure was transformed to the SAO basis and atomic contributions attributed as follows (see Fig.~\ref{fig:SAO decomp bands}): We first found the peak height percentages $p^\mathrm{SAO}(E)$ for high symmetry points along the bands near the Fermi level. Then the total peak heights can be attributed to all atomic orbitals (AO) via the normalized coefficient vectors $C^\mathrm{AO}$ as
\begin{equation}
    p_{i}^\mathrm{AO}(\vec k,E) = \sum_j p_{j}^\mathrm{SAO}(\vec k,E) [C_i^\mathrm{AO}(j)]^2,
\end{equation}
where $i$ and $j$ are indices for AO and SAO respectively, and $\vec k$ is the coordinate of high symmetry point in the reciprocal space. Results are shown in Fig.~\ref{fig:AO decomp bands}. The AO percentages at a fixed $\vec k$ and $E$ are normalized as
\begin{equation}
    \sum_ip_{i}^\mathrm{AO}(\vec k,E) = 1.
\end{equation}

\textit{Electronic density.} In order to simulate the electronic density on the $xy$-plane, we used the four dominant SAOs shown in Fig.~\ref{fig:SAO decomp bands} that contributed to the bands at $\Gamma$ near the Fermi level. The partial occupation of each SAO was determined by solving the interpolated density matrix after orthogonalization. Then the density was scaled by its ratio to the total occupation for these four SAOs. The combined electronic density was plotted in Figs.~\ref{fig:Fig4}(d) and (e). The unit cell and its electronic density are repeated along the $x$ and $y$ direction.

\end{document}